\documentclass[twocolumn,amsmath,10pt,showkeys,floatfix]{revtex4-1}
\pdfoutput=1
\usepackage[latin1]{inputenc}
\usepackage{color} 
\definecolor{darkblue}{rgb}{0,0,0.5}
\definecolor{lila}{rgb}{0.3,0,0.3}
\definecolor{turq}{rgb}{0,0.1,0.4}
\definecolor{lightblue}{rgb}{0.7,0.7,0.9}
\usepackage{graphicx}
\usepackage{enumerate}
\usepackage[pdftex,
  colorlinks=true,
  backref=page,
  linkcolor=darkblue, 
  filecolor=red,
  citecolor=turq, 
  urlcolor=lila, 
  pdftitle={Homomorphic Payment Addresses and the Pay-to-Contract Protocol},
  pdfauthor={Ilja Gerhardt, Timo Hanke},
  pdfsubject={Homomorphic Payment Addresses and the Pay-to-Contract Protocol},
  pdfkeywords={financial cryptography, online electronic payment protocol, payment scheme, homomorphic keypairs, ElGamal, bitcoin, deterministic wallet, labeled wallet, offline anonymous merchant},
  pdfpagelabels=true, 
  breaklinks=false,
  plainpages=false,
  backref=false,
  bookmarks,
  bookmarksnumbered=true]{hyperref}

\newcommand{\ci}[1]{\large\textcircled{\small{\texttt{#1}}}\normalsize}

\begin{document}

\title{\large{Homomorphic Payment Addresses and the Pay-to-Contract Protocol}}
\author{Ilja Gerhardt}
\email{ilja@quantumlah.org}
\affiliation{Max Planck Institute for Solid State Research,
  Heisenbergstraße 1, D-70569 Stuttgart, Germany}

\author{Timo Hanke}
\email{hanke@math.rwth-aachen.de}
\affiliation{Lehrstuhl D f\"ur Mathematik, RWTH Aachen, Templergraben 64, D-52062 Aachen, Germany}

\date{\today}

\begin{abstract}
We propose an electronic payment protocol for typical customer-merchant relations 
which does not require a trusted (signed) payment descriptor to be sent from the merchant to the customer.
Instead, the destination ``account'' number for the payment is solely created on the customer side.
This eliminates the need for any encrypted or authenticated communication in the protocol
and is secure even if the merchant's online infrastructure is compromised.
Moreover, the payment transaction itself serves as a timestamped
receipt for the customer.
It proves what has been paid for and who received the funds,
again without relying on any merchant signatures.
In particular, funds and receipt are exchanged in a single atomic action.
The asymmetric nature of the customer-merchant relation is crucial.

The protocol is specifically designed with bitcoin in mind as the underlying payment system.
Thereby, it has the useful benefit that all transactions are public.
However, the only essential requirement on the payment system is that ``accounts'' are arbitrary user-created keypairs of a cryptosystem
whose keypairs enjoy a homomorphic property.
All ElGamal-type cryptosystems have this feature.
For use with bitcoin we propose the design of a deterministic bitcoin
wallet whose addresses can be indexed by clear text strings.
\end{abstract}


\maketitle

\section{Introduction}

We first review the current best-practice for bitcoin payments in customer--merchant relations,
discuss its obvious shortcomings under a reasonable threat model,
and formulate a goal for improvement.

\subsection{Bitcoin}

The Bitcoin network was introduced as a peer-to-peer online currency in 2009~\cite{nakamoto}.
To prevent double spending, a peer-to-peer currency faces the problem of establishing mutual agreement across the entire network regarding the ownership of funds.
Bitcoin solves this fundamental problem by a so-called {\em proof-of-work} principle, 
which requires every node in the network to solve a computationally hard challenge before 
that node's ownership data is accepted by other peers.
This challenge is defined by bitcoin as partially inverting the cryptographic hash function SHA-256.
Mutual agreement regarding ownership is recorded in an authenticated data structure called the {\em blockchain}. 
Data blocks in the chain record ownership updates (called {\em transactions}) rather than balance sheets.
Each transaction is signed against the public key of the previous owner
(the {\em input} of the transaction)
and defines the public key of the new owner
(the {\em output}).
In fact, the bitcoin system only deals with keypairs
--- the ``owner''  of a keypair becomes an irrelevant entity and is in no way recorded.
Consequently, bitcoin does not rely on any public key infrastructure (PKI).
One commonly refers to public keys as {\em addresses},
suggesting that they are the ``location'' of funds.
Users hold collections of keypairs, and such collections are called {\em wallets}.
Currently, all keypairs for bitcoin belong to the ECDSA signature scheme (Elliptic Curve Digital Signature Algorithm)
on the elliptic curve secp256k1~\cite{secg}.

This paper uses a homomorphic feature of keypairs, that ECDSA shares with all ElGamal-type signature schemes 
(see end of~\ref{ssec:det} below for details)
and would not be possible if bitcoin used RSA-type signatures.
Conversely, bitcoin can be replaced in this paper by an abstract payment system 
that allows arbitrary user-generated account numbers 
which correspond to public keys of a cryptosystem with equivalent or similar homomorphic features.   
For instance, the decentralized aspect and the proof-of-work principle of bitcoin play no role here.
Other details of the bitcoin technology such as the {\em script system} are not essential for the main part of the paper 
but are addressed in the dedicated subsection~\ref{ssec:script}. 

\subsection{The customer-merchant relation}

A customer $C$ has a device $D$ (e.g.\ a desktop computer) to browse merchant websites and to place orders on those sites.
$D$ does not hold any bitcoin wallet files.
Furthermore, $C$ interacts with a trusted path device $T$ (e.g.\ a
hardware wallet) to sign transactions. 
For this purpose, $T$ has access to the private keys of a bitcoin wallet.
We assume $T$ is never compromised.
This assumption is justified, e.g., in a scenario where $T$ is securely booted, receives data from $D$ through an air gap,
and has a one-way channel to broadcast signed transactions to the bitcoin network.

\begin{figure}[h]\label{fig:unsecure}
  \centering
  \includegraphics[width=\columnwidth]{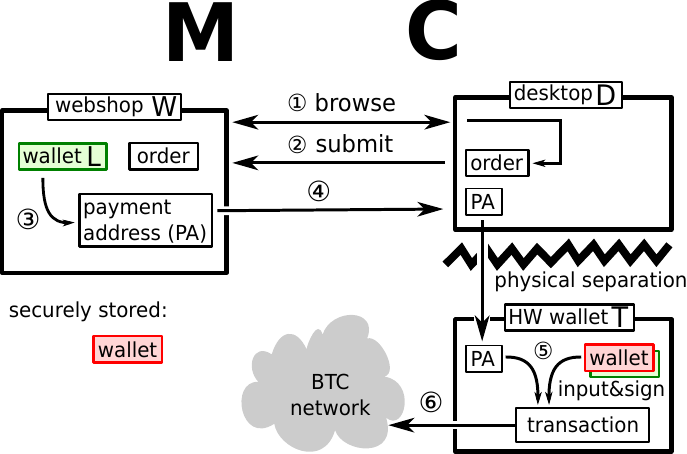} 
  \caption{Classic protocol for bitcoin payments in customer-merchant relations. 
\ci{1} customer $C$ with his desktop computer $D$ browses the webshop $W$ run by merchant $M$ over a secure SSL connection.
\ci{2} $C$ submits an order to the webshop.
\ci{3} $W$ picks a payment address from a pool $L$ of bitcoin addresses,
which comprises the public part of a bitcoin wallet (green). 
The private part of the wallet (red) is stored offline.
\ci{4} The payment address is returned to $D$.
\ci{5} $C$ signs a transaction to the payment address on his hardware wallet $T$.
\ci{6} $C$ broadcasts the transaction to the bitcoin network.}
\end{figure}

A merchant $M$ has a device $W$, e.g.\ a server running a webshop, to receive and process orders.
Processing an order requires at least to generate a payment address and to return it to the customer.
For this purpose,
$M$ divides his bitcoin wallet in two parts, the {\em address wallet} $L$ and the {\em key wallet}.
The address wallet is stored on $W$, whereas the key wallet is stored securely offline.
The commonly implemented payment protocol for this standard scenario is depicted in Figure~\ref{fig:unsecure}.

\subsection{Signatures and merchant reputation}

We first focus on security of the merchant side.
As a reasonable threat model, since $W$ is connected to the Internet and is a publicly known webshop,
we assume that $W$ is compromised by an attacker $A_1$.
Since the key wallet is offline, $A_1$ cannot steal coins directly from $M$.
However, $A_1$ can tamper with the payment address and redirect $C$'s transaction to his own addresses.
Depending on the type of order,
$A_1$ may also tamper with other information in the order, e.g.\ a delivery address, and redirect goods instead of funds.

This attack can -- in principle -- be prevented by introducing signatures into the protocol.
We assume that $M$ has a private/public keypair $(K,P)$ for which the public key $P$ is known to $C$ before the protocol starts.
An obvious choice for $(K,P)$ would be the keypair on which $M$ builds his reputation as a merchant, e.g.\ a Secure Sockets Layer (SSL) certificate.
Conversely, however, we do not assume that $M$ holds a public key of $C$ before the protocol starts.

Suppose $M$ bundles the order (which we assume contains the delivery address) together with a payment address into a {\em bill},
signs the bill with $K$ and returns the signed bill to $C$.
For $C$ this means that, if anything goes wrong, he can prove that the attack was launched on the merchant side and not on his side.
For $M$, however, this means that signing the bill can not be done on $W$.
The naive idea that $M$ signs all addresses from the address wallet $L$ offline 
and stores the pre-signed version of $L$ on $W$ is not sufficient.
While it protects the funds well, it does not protect the goods
since there is no way to pre-sign the delivery addresses.
Generating addresses from so-called deterministic wallets (see~\ref{ssec:det} for a definition) also does not help.

\subsection{The firewall solution}

\begin{figure}[h]
  \centering
  \includegraphics[width=\columnwidth]{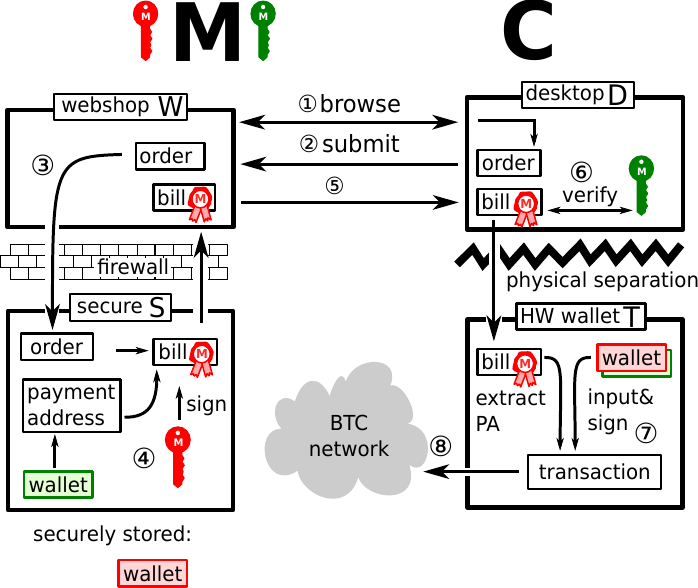} 
  \caption{Signature-based protocol with firewall.
\ci{1},\ci{2} as before.
\ci{3} $W$ passes the order across a firewall to device $S$.
\ci{4} Order and payment address are signed {\em together} with the private key (red key) of $M$, forming the bill.
\ci{5} The bill is returned to $D$.
\ci{6} $D$ verifies the bill's signature against the public reputation key (green key) of $M$.
\ci{7} $C$ signs a transaction to the bill's payment address on his hardware wallet $T$.
\ci{8} $C$ broadcasts the transaction to the network.}
\label{fig:firewall}
\end{figure}

In the ``firewall solution'' $M$ stores the address wallet $L$ and the private reputation key $K$ on a separate, well-protected device $S$.
We assume that $S$ is linked to $W$ via a channel that does not allow $A_1$ to compromise $S$ (firewall).
After receiving and pre-processing the order, $W$ forwards the order to $S$.
Then $S$ picks an address from $L$, signs address and order together into a bill,
sends the signed bill to $W$ who forwards it to $D$.
For this purpose, 
$S$ instead of $W$ holds $L$ and $K$.
This protocol is depicted in Figure~\ref{fig:firewall}.

While it is feasible for $C$ to interact with a trusted path device,
$M$ wants $S$ to run unattended.
This raises two problems:
The first problem is to implement a secure link between $W$ and $S$.
This may set off individuals and small businesses who do not
have the required expertise.  
A second problem is to properly protect $S$ physically.
This may set off even mid-sized businesses who, on the one hand,
have to replicate $S$ in order to scale or to prevent denial-of-service attacks,
and, on the other side, 
do not have the resources to protect all replications of $S$.

\subsection{Customer side security}

We now turn to the security of the customer side,
assuming that the device $D$ is compromised by an attacker $A_2$.
Then $A_2$ can play a man-in-the-middle attack between $M$ and $C$,
e.g.\ submitting a false delivery address to $M$.
Or, $A_2$ can forward a false payment address to $T$.
The only solution is to carry out the verification of the signed bill against the merchant's public key $P$ on the trusted device $T$ rather than on $D$.
In order to make this verification reliable,
we assume that $P$ is stored on $T$ {\em before} $C$ places an order.
However, even though $T$ can verify the bill's signature automatically,
this does not mean that the whole verification process is automatic.
Rather, in order to prevent $A_2$'s man-in-the-middle attack,
$T$ has to display the whole bill and wait for an interactive confirmation from the customer $C$.

We remark that if $D$ is compromised by $A_2$ then $C$ cannot rely on $D$ to authenticate $W$,
e.g.\ via the usual SSL/TLS authentication of the browser running on $D$.
However, having the signature verification against $P$ take place on $T$,
this does not pose a problem.
In fact, authentication between $W$ and $D$ is not necessary anymore.

The assumption that $P$ is stored on $T$ before the protocol starts is fundamental throughout the paper. 
Public key infrastructures (PKI) can solve the problem how $C$ obtains $P$.
A possible scenario is that hardware wallets are shipped with hard-coded root keys, just like browsers are shipped with root SSL/TLS certificates. 
This, however, is not in the scope of this paper.

\subsection{Goal}\label{ssec:goal}

The goal is to devise a payment protocol that works without an additional secure device $S$ on the side of the merchant.
Since $M$ cannot store the private reputation key $K$ on $W$, the protocol must be {\em signature-less}.
This can only be achieved by moving the generation of the payment address to the {\em customer side}.

The idea for such a protocol is not new.
Customer-generated payment addresses are suggested in \cite{andresen},
and \cite{lerner} points out that they dispense with the need of encrypted or authenticated communication between customer and merchant.

\subsection{Anonymity}

We also discuss anonymity properties of our payment protocol.
However, we disregard all issues arising from tracing payments via an analysis of the public transaction record.
The same holds for other (non-payment) order-related information,
that may be obtained via tracing shipments or downloads.
In particular, it is not in the scope of this paper
how the customer can retain his anonymity from the merchant.

Just like with any protocol that creates a new payment address for every transaction,
one can easily conceal that $M$ is the addressee of a payment transaction (see \ref{ssec:summary}).
We discuss in further detail
how the merchant can hide himself and all his infrastructure
from the customer (section \ref{sec:anon}).
This requires the merchant to operate under a pseudonym,
which we assume is represented by the reputation key $P$. 

\section{The pay-to-contract protocol}

We first discuss the customers requirements regarding receipts in~\ref{ssec:receipts}
and formulate the {\em pay-to-contract principle}.
Then, in~\ref{ssec:label}, we devise a protocol adhering to this principle that does not rely on any form of trusted (signed) payment descriptor.

\subsection{Payment receipts}\label{ssec:receipts}

After making a payment, $C$ wants to be able to prove 
\begin{enumerate}[(i)]
\item that $M$ has agreed to fulfill a given order upon receiving a certain amount of funds,
\item that a certain amount was paid to and received by $M$ for that same order. 
\end{enumerate}
In the traditional, paper based customer--merchant relation (i) is accomplished by a {\em bill}, signed by $M$ in advance,
and (ii) by a {\em receipt}, signed by $M$ in exchange for the funds.
The receipt usually contains a reference to the bill.
In fact, for an electronic payment protocol we simply assume the
receipt contains the bill --- just as paper based receipts are usually
kept stapled together with the corresponding bill by the customer.

More abstractly, the bill is a {\em contract} $x$ signed by $M$ alone, 
to which $C$ enters by conducting the payment required in $x$.
If a cryptographic hash of $x$ can be attached to the transaction on the blockchain
then this transaction proves (ii).
There are several known ways to attach a unique hash to a bitcoin transaction,
e.g.\ by encoding the hash into one of the output addresses.
We will systematize such an encoding in~\ref{ssec:label} below.  
Thus, with bitcoin, funds and receipt can be exchanged in an atomic action.
Since the blockchain is public, the contract $x$ proves both (i) and (ii),
i.e.\ functions as bill {\em and} receipt. 

\subsection{Pay-to-contract}\label{ssec:p2c}

A {\em payment descriptor} for a contract $x$ specifies the payment that $C$ has to conduct in order to enter $x$.
It is a record, holding a payment address and an amount, that is signed by $M$ with a reference to $x$.
Alternatively, one can think of the payment descriptor as part of $x$ and require that $x$ is signed by $M$ as a whole.
 
The main point of this proposal is to eliminate payment addresses from all payment descriptors,
whether they are part of $x$ or separate.
Instead, we suggest that the payment address is derived deterministically from the contract itself,
so that it can be computed by $C$ alone and without any additional data.
This is what we call the {\em pay-to-contract principle} and what achieves the goal formulated in~\ref{ssec:goal}. 

In detail, pay-to-contract works as follows:
As in the setup of the introduction, $K$ is the private key and $P$ the public key 
that define the online reputation of $M$.  
Let $H$ be a cryptographic hash function.
Suppose we have a pair of functions $d_{\text{addr}},d_{\text{priv}}$ such that
$d_{\text{addr}}(P,H(x))$ is a bitcoin address unique for $P$ and $H(x)$, 
$d_{\text{priv}}(K,x)$ is the private key corresponding to $d_{\text{addr}}(P,x)$,
and $d_{\text{priv}}(K,x)$ cannot be computed without $K$. 
If we define $b:=d_{\text{addr}}(P,H(x))$ as the payment address for $x$,
then $C$ can compute $b$ with the information available to him
and only $M$ can access the funds on $b$.

As an additional feature, $x$ automatically serves as a receipt for $C$ in the sense of~\ref{ssec:receipts}. 
Since $H(x)$ is encoded in the output address $b$, the contract $x$ is verifiably attached to the payment transaction\footnote{Specifically for bitcoin, we remark that there is no need to include $H(x)$ {\em explicitly} into the transaction with
mechanisms such as an \texttt{OP\_DROP} operation in the {\em scriptsig}.}.

A suitable {\em address derivation function} $d_{\text{addr}}$ can be obtained from deterministic wallets,
as was previously noted in \cite{andresen}, \cite{lerner}.
Before elaborating on the details (\ref{ssec:protocol}),
we recall the concept of a deterministic wallet (\ref{ssec:det})
and propose to standardize a certain type of deterministic wallet (\ref{ssec:label}).

\subsection{Deterministic Wallets}\label{ssec:det}

A {\em deterministic wallet} is a sequence of bitcoin keypairs (i.e.\ addresses and their private keys)
that are all derived from a single secret called the {\em base} of the wallet.
For instance, the base can be a bitcoin keypair by itself, 
but this is not a requirement. 

Deterministic wallets greatly facilitate backups because it suffices to back up the base.
All keypairs can be recovered from the base by enumerating them.

One further distinguishes two types of deterministic wallets~\cite{gmaxwell}:
In a {\em type 2} wallet, one can derive from the base another secret that we call the {\em pubbase}.
All addresses can then be derived from the pubbase alone, but none of the private keys can.
{\em Type 1} wallets do not have this feature.
All known type 2 wallets have the disadvantage that the base can be reconstructed from the pubbase and any one of the wallet's private keys,
thus diminishing somewhat the usefulness of the pubbase.
For type 1 wallets it is trivial to prevent this,
i.e.\ to make the security of any of the wallet's private key independent from the security of the base.

The separation of base and pubbase in type 2 wallets offers two significant benefits.
First, there is a security benefit:
a vulnerable device can store only the pubbase and derive addresses as they are needed,
while the private keys can be derived independently on another device that holds the base.
Second, it is possible to prove that a certain pubkey was derived from
a given pubbase without revealing any private keys. 

Type 2 deterministic wallets are possible because from an ElGamal-type  keypair $(s,a)$ 
one can obtain a new keypair $(ns,na)$ by multiplication with any integer $n$
(see~\cite{elgamal} or sections 8.4, 11.5.2 in \cite{menezes}).
This is called a homomorphic relation between private and public keys. 
Type 2 deterministic wallets would not be so easily available if bitcoin used RSA signatures, for instance.
 
\subsection{Labeled Wallets}\label{ssec:label}

By a {\em labeled wallet} we mean a (non-enumerable) set of bitcoin keypairs 
that are all derived from a single secret called the {\em base} of the wallet 
and that are indexed (relative to the base) by an arbitrary {\em label}.

The difference between a labeled wallet and a deterministic wallet is that labels cannot be enumerated.
To backup a labeled wallet it does not suffice to backup the base,
one also needs to know the labels.

Labeled wallets can mimic deterministic wallets by restricting the labels to some enumerated sequence.
Conversely, a deterministic wallet in which the enumerated keypairs
can be accessed randomly can mimic labeled wallets by translating the
label to a number via a hash function. 

The same distinction between type 1 and 2 can be made for labeled wallets as for deterministic wallets.
We propose in the following to standardize a type 2 labeled wallet.
To this end, we first fix
i) an encoding for the labels (e.g.\ UTF-8),
and ii) a cryptographic hash function $H$ (e.g.\ SHA-256) whose digests are interpreted as non-negative integers.
The application of labeled wallets may not be restricted to the pay-to-contract principal.

As a {\em base} for the labeled wallet we use an ECDSA keypair $(s,a)$ for bitcoin's curve secp256k1~\cite{secg},
i.e.\ $s$ is an exponent and $a$ is the point $g^s$ where $g$ is the base point of the curve.
We call $s$ the {\em private base} (or {\em master private key}) 
and $a$ the {\em pubbase} (or {\em master pubkey}).
The curve operation is written multiplicatively in order to better distinguish points from exponents (other people prefer to write $a=s\cdot g$ instead).

The {\em derived keypair} with label $x$ of the wallet with base $(s,a)$ is written $(s[x],a[x])$ 
and is constructed as follows. 
The label is an arbitrary string in the chosen encoding (e.g.\ $x=$'savings1').
We define
\begin{equation}
  \label{eq:derkey}
 s[x]:=s+H(x), \quad a[x]:=a\cdot g^{H(x)}.
\end{equation}
We call $s[x]$ the {\em derived private key} and $a[x]$ the {\em derived pubkey}.
Obviously, the derived pubkey can be computed from $a$ and $x$ without knowledge of $s$.

In order to apply labeled wallets to the pay-to-contract principle,
we suppose that $M$'s keypair $(K,P)$ is an ECDSA keypair for the same curve secp256k1.
Then $(K,P)$ can serve simultaneously as the base of a labeled wallet.
In effect, $M$ builds online reputation {\em on a labeled wallet}.
Functions $d_{\text{addr}},d_{\text{priv}}$ that satisfy the requirements of~\ref{ssec:p2c}
are obtained by 
\begin{equation}
  \label{eq:f}
  d_{\text{priv}}(P,x):=K[x], \quad d_{\text{addr}}(P,x):=\text{Hash160}(P[x]),
\end{equation}
where Hash160 is the bitcoin-defined 20-byte hash.

\subsection{Basic protocol}\label{ssec:protocol}

A pay-to-contract protocol using labeled wallets and address derivation \eqref{eq:f} works as follows.
Upon receiving an order from $C$'s device $D$, device $W$ generates a contract $x$ and sends $x$ back to $D$.

The part of $x$ that does not change on a order-by-order basis is called {\em static}.
Static information may change in regular intervals that are convenient to $M$,
usually spanning several orders.
For instance, $M$'s terms of service and price list would be considered static.
The part of $x$ that changes on a order-by-order basis is called {\em dynamic}.
All information given by $D$ in an order,
e.g.\ the item number, quantity and delivery address, 
is considered dynamic.
 
\begin{figure}[h]
  \centering
  \includegraphics[width=\columnwidth]{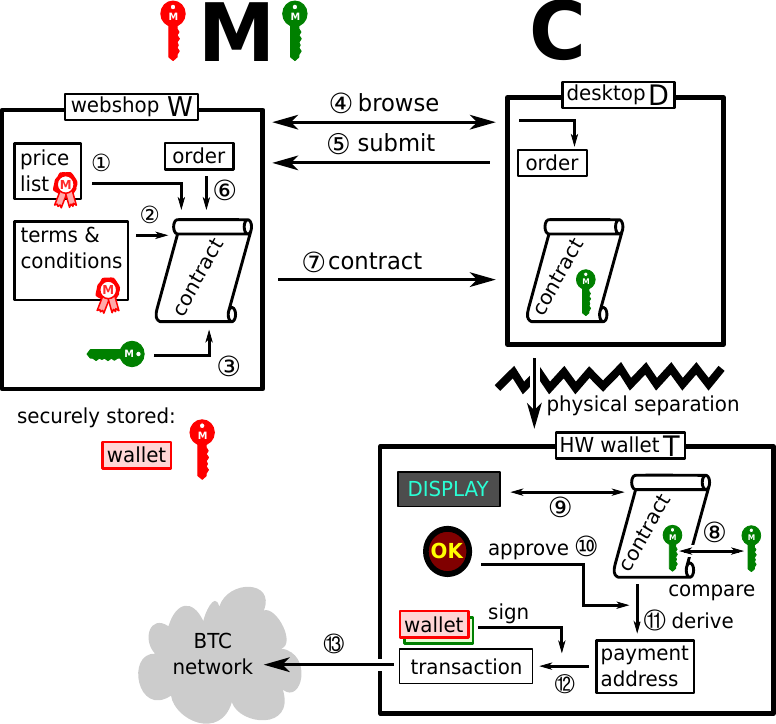} 
  \caption{Pay-to-Contract protocol.
\ci{1}--\ci{3} Merchant $M$ signs price list and terms and conditions offline. 
Pre-signed versions plus $M$'s public reputation key are compiled into contract template, stored on webshop $W$.
\ci{4}--\ci{5} as before.
\ci{6} $W$ fills order details into contract template, forms contract.
\ci{7} Contract returned to $D$.
\ci{8} $C$ uses hardware wallet $T$ to verify contract signatures.
Contract's reputation key is looked up in pre-stored key--alias database.
\ci{9} $T$ displays {\em all} contract details plus alias of reputation key.
\ci{10} $C$ manually approves contract details and identity of reputation key.
\ci{11} $T$ applies address derivation function to contract, obtains payment address.
\ci{12}--\ci{13} as before.}
\label{fig:p2contract}
\end{figure}

We assume that $x$ contains the pubkey $P$,
all static information is signed with $K$,
and dynamic information is unsigned.
Upon receiving $x$, device $D$ checks if $x$ is consistent with the order it intended to place,
and, if so, forwards $x$ to $T$.
Then $T$ verifies all signatures inside $x$ and displays $P$ and $x$ to $C$.
When $C$ interactively approves the contract, $T$ computes $b:=d_{\text{addr}}(P,x)$ and generates and broadcasts a transaction that sends the amount given in $x$ to address $b$. 
Finally, $T$ permanently stores $x$ because the contract is
simultaneously the receipt for $C$.
This protocol is depicted in Figure~\ref{fig:p2contract}.
\subsection{Security}

Since $C$ only approves a transaction if $T$ displays an order $x$ that $C$ actually intends to place,
the transaction will always go to the correct output address $d_{\text{addr}}(P,x)$.
If an attack leaves $W$ with a tampered version $x'$ of $x$
then $M$ expects the payment at a different address $d_{\text{addr}}(P,x')$.
The result will be that $M$ never detects a payment to this address and never ships the order.
When $C$ submits the receipt $x$ to $M$ via another channel to resolve the matter (even publicly, in court),
then $M$ but only $M$ can compute $s[x]$ and redeem the output.

Therefore, the attack possibilities of $A_1,A_2$ are limited to denial-of-service 
and messing with the tracking and inventory system that may run on $W$.
It cannot be proved whether the attack happened on the customer or merchant side.

\subsection{Summary of properties}\label{ssec:summary}

A few properties of the pay-to-contract protocol are worth being summarized:
\begin{enumerate}
\item There is a unique address for each contract $x$ and a public proof of payment.
Anyone who knows $x$ can verify whether $x$ was paid for or not.
$M$ cannot deny to have received a payment to $d_{\text{addr}}(P,x)$.
\item Third parties cannot necessarily monitor all incoming payments to $M$, because they would have to guess $x$.
If $x$ is properly randomized, e.g.\ if $x$ contains a salt, 
then neither $P[x]$ nor $d_{\text{addr}}(P,x)$ can be linked to the merchant via his public key $P$.
\item 
The major part of the protocol can be carried out on untrusted devices ($D$ and $W$).
In particular, the communication link between $D$ and $W$ is not required to be secure (encrypted or authenticated).
Instead, it can be unencrypted email.
\item
$C$ needs a trusted path device $T$ with 
a display large enough to show all sensitive parts of the contract~$x$. 
In particular, it cannot be a smart card, and single-line LCD displays also seem improper.
\item 
$C$ must have obtained $P$ in advance,
and must have stored $P$ in a place where no attacker can tamper with it
and from where $C$ is able to compare $P$ to what $T$ displays.
Of course, $P$ can be stored on $T$ itself, e.g.\ linked to an alias that is displayed instead of $T$.
In this case, the binding of $P$ to an alias can further be authenticated by a public key infrastructure
(including web-of-trust).
\item\label{it:offline}
For repeated orders, $C$ can create $x$ all by himself with no direct interaction with $M$.
Only after the transaction does $C$ need to submit $x$ to $M$.
This feature is further used and explained in~\ref{sec:anon} below.
\end{enumerate}

Authentication in this protocol is moved from the browser level (based on trust in SSL/TLS certificates)
to the payment level (based on the trust in the key $P$).
We remark that building online reputation (i.e.\ reputation as a merchant) on a ``payment base address'' (such as $P$) is more universal than on SSL/TLS certificates.
For example, 
a business may have a single payment base address but several websites under different top-level domains
who would all require their own SSL/TLS certificate. 
Moreover, certain scenarios (e.g.\ paying at point-of-sales, teller machines, etc.) may not involve any browser at all,
leaving no other choice than to authenticate the payment address directly.

\subsection{Generalized Labeled Wallets}\label{ssec:script}

We now discuss a bitcoin specific enhancement based on the so-called script system.
In bitcoin, transaction outputs are more flexible than simple ``account numbers''.
Instead, they are stack-based scripts.
An output can only be redeemed upon providing an input to its corresponding script such that, when run with this input, the script returns true.
Due to the significantly improved security it provides,
the so-called ``multi-signature'' script is becoming popular.
It requires signatures for a threshold number of $n$ public keys from a set of $m$ specified public keys.
Merchants likely want to receive funds directly to multi-signature scripts, rather than to regular bitcoin addresses.

Complicated scripts are often replaced by their hash in a special type of output called {\em pay-to-script-hash} (P2SH, ~\cite{p2sh}).
This can be done with any script and does not affect the enhancement we are about to describe.

As it turns out, the pay-to-contract protocol can be generalized in a straightforward manner to produce a wide array of script outputs.
The idea is to define a {\em base script} in place of a base address.
From this, one obtains {\em derived scripts} which can only be spent with the same knowledge required to spend the base script.
The derivation process for label $x$ replaces every pubkey $P$ in the base script by the derived pubkey $P[x]$
in the sense of~\ref{ssec:label}.
In this way, all scripts that contain explicit pubkeys and not their hashes can serve as base scripts.

As an example, we look at a 2-of-2 multi-signature script:
\begin{equation}
  \label{eq:multisig}
\texttt{2 $P_1$ $P_2$ 2 OP\_CHECKMULTISIG} 
\end{equation}
where $P_1,P_2$ are pubkeys.
Using \eqref{eq:multisig} as the base script, the derived script with label $x$ is obtained by replacing each pubkey $P_i$ with $P_i[x]$:
\begin{equation}
\label{eq:der-multisig}
\texttt{2 $P_1[x]$ $P_2[x]$ 2 OP\_CHECKMULTISIG} 
\end{equation}
For practical reasons, the customer then pays to the script hash address (P2SH) of \eqref{eq:der-multisig}, i.e.\ to:
\[ \texttt{OP\_HASH160 <addr> OP\_EQUAL} \]
where \texttt{<addr>} is the 20-byte hash of \eqref{eq:der-multisig}.

If the merchant wallet ``base'' is a script rather than a pubkey 
then it can obviously no longer be identical to the reputation key $P$.
Instead, it should be signed by the reputation key and publishes via some public key infrastructure.

\subsection{Alternative definitions}

Alternatively to \eqref{eq:derkey}, we can set
$s[x]:=s\cdot H(x)$ and $a[x]:=a^{H(x)}$ without
changing any properties of the labeled wallet
nor any properties of the protocol.

\section{Structure of contracts}
\subsection{Contract as Merkle trees}
In the pay-to-contract protocol, the contract $x$ serves simultaneously as the payment descriptor and the receipt.
Its function as a receipt demands that $C$ is able to partially reveal the contents of $x$.
In the extreme case, $C$ may only want to prove that a certain transaction made by him has an output that is controlled by $M$,
while in other cases $C$ may want to reveal what was paid for but not the delivery address.
Therefore, $x$ is structured in a Merkle tree and every node is salted.
In order to hide the contents of a branch,
it is replaced by its hash digest without making $x$ invalid.  

The layout of the tree does not need to be standardized,
its design can be left to $M$.
However, we propose that at least the overall format of a contract as a JSON-encoded tree-object is standardized.

Leaf nodes of the tree can further be encrypted with a public key that is specified in the contract layout.
In this case, the hash of the ciphertext (not the cleartext) is used in the Merkle tree.
Then $C$ has the choice to reveal the cleartext, the ciphertext or the hash.
As an application of encrypted contract fields,
one can think of an entity that is handling contracts on behalf of $M$.
This entity can receive the contract with encrypted fields, can verify that is has been paid for on the blockchain,
and can pass it on to $M$, all without seeing the cleartext of the encrypted fields.

\subsection{Recommended fields of contracts}

Even though the tree layout is left to the merchant,
it is advisable to make recommendations for the naming of fields,
in order to facilitate automatic verification.
This can include a division into {\em static} and {\em dynamic} fields,
depending on whether the field is changed on an order-by-order basis or not.
It is recommended that static fields are signed with the merchant's online reputation,
whereas dynamic fields are signed with another key that represents the merchant's online tracking and inventory system.

A suggested list of dynamic data records is:
item number,
quantity,
price,
delivery address,
delivery deadline (timestamp or block number),
payment address,
payment deadline (timestamp or block number),
client's reference/tracking token,
merchant's reference/tracking token. 

A preliminary list of static data records is:
price,
merchant's terms of service,
warranty,
name of merchant's auditors.
Static data records should be pre-signed by the merchant before they are assembled into the contract. 

\section{Offline and anonymous merchants}\label{sec:anon}

\subsection{Offline merchants}\label{ssec:offline}

\begin{figure}[h]
  \centering
  \includegraphics[width=\columnwidth]{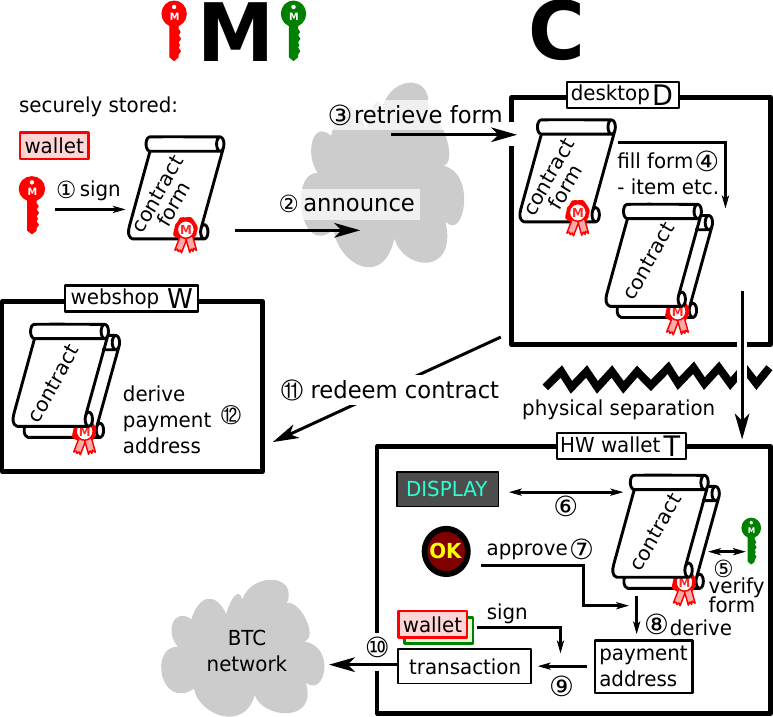} 
  \caption{Offline protocol
\ci{1} Merchant $M$ signs a {\em contract form} offline, containing price list, terms and conditions, expiration date.
\ci{2} $M$ posts contract form in a public place (electronic or print).
\ci{3} Customer $C$ retrieves contract form from public place.
\ci{4} $C$ fills order details (item, quantity, delivery address) into contract form, forming a contract.
\ci{5}--\ci{10} as before. 
\ci{11} $C$ redeems contract to webshop $W$ via one-way channel.
\ci{12} $T$ applies address derivation function to contract, obtains payment address to monitor.}
\label{fig:offline}
\end{figure}

According to~\ref{ssec:summary}, point~\ref{it:offline}, the customer can order ``offline'' and pay in advance
before informing $M$ about it.
For instance,
if $M$ sells digital (unlimited) goods for a publicly known price
or simply receives donations
then $C$ has no need to communicate with $M$ before the payment.
Neither the order itself nor the pay-to-contract protocol require such communication.
After the payment, $C$ can redeem the contract $x$ for the digital goods.
This protocol is depicted in Figure~\ref{fig:offline}.
It requires a public place for the announcements of ``blank''
contracts called {\em contract forms}.
For practical purposes, all announcements from the merchant $M$ should be tagged by
the public reputation key of $M$. 
 
The situation is different with physical goods because usually $C$ wants to check with $M$ first if the items are in stock.
 
Granted that $C$ knows $M$'s reputation pubkey in advance, 
the interaction between $M$ and $C$ in Figure~\ref{fig:offline} 
is reduced to a {\em one-way communication} from $C$ to $M$ (the redemption of $x$).
This means instead of running an interactive webshop,
$M$ only needs a channel to receive contracts (e.g.\ email).
Going even further, customers can post their (encrypted) contracts
in a public place from where $M$ retrieves them.
In this case there is no direct communication between $C$ and $M$ at all,
neither for ordering nor for paying,
and we speak of $M$ as an {\em offline merchant}.

Being an offline merchant may help $M$ to stay anonymous from $C$, 
or may make $M$ less prone to denial-of-service attacks.

\subsection{Anonymity}

An offline merchant $M$ may want to stay anonymous and use his reputation key as a pseudonym.
Moreover, $M$ may want to hide as much of his activity from the public as possible.
He therefore asks his customers to redeem their contracts by posting them in encrypted form to a public distributed filesystem. 

Then $M$ is required to monitor the filesystem and to identify the contracts posted for him among all files.
We assume it is infeasible for $M$ to download and check every single file whether he can decrypt it or not. 
Therefore, the customers have to ``tag'' their posted files in some way as being addressed to $M$.
But this is not easily possible without revealing to the public that a certain file is addressed to $M$.
Hence, naively using the above offline protocol may provide anonymity to $M$ but leaks information such as the number of orders that $M$ receives, their timing and contract sizes to any observer of the distributed filesystem. 

\subsection{Goal and Achievement}

The goal is to devise a protocol that allows customers of
an offline merchant to redeem contracts without disclosing to an observing third party
that a redemption has taken place.
If this is achieved we speak of $M$ as a fully {\em anonymous merchant}.

\begin{figure*}
  \def\@captype{figure}
  \centering
  \includegraphics[width=1.5\columnwidth]{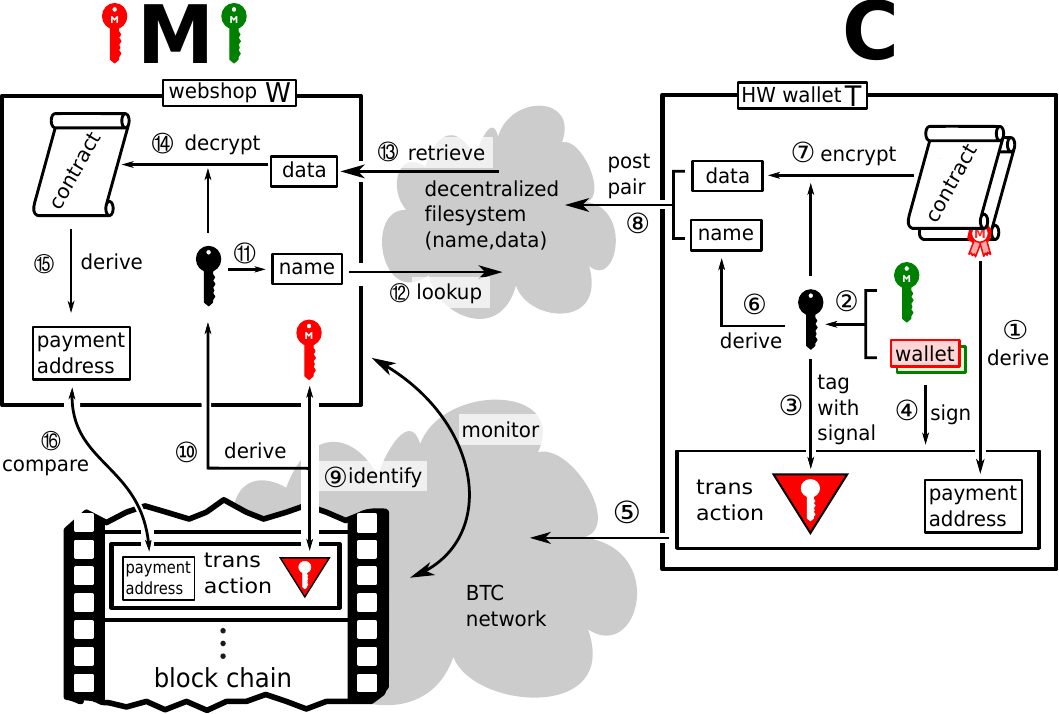} 
  \caption{Payment and signaling in one transaction.
Suppose \ci{1}--\ci{7} of Figure~\ref{fig:offline} are carried out,
these steps are not shown here.
\ci{1} $T$ applies address derivation function to contract, obtains payment address.
\ci{2} Diffie-Hellman key agreement produces shared secret (black key).
\ci{3} Transaction is secretly tagged with shared secret.
\ci{4}--\ci{5} Transaction is signed and broadcast to the network. 
\ci{6} A filename is derived from shared secret. 
\ci{7} Contract is encrypted with shared secret as symmetric key.
\ci{8} Ciphertext is posted to distributed filesystem under derived filename. 
\ci{9} $W$ monitors the blockchain and identifies tagged transactions.
\ci{10} Shared secret is derived from tag.
\ci{11}--\ci{12} Filename is derived from tag and looked up.
\ci{13}--\ci{14} Data is retrieved and decrypted, $W$ obtains contract.
\ci{15} $W$ derives payment address is derived.
\ci{16} Order is accepted if payment address equals transaction output.
}
\label{fig:anon}
\end{figure*}

We divide the protocol into two parts.
First, the {\em signaling protocol} allows $C$ to secretly share a unique integer value with $M$. 
By this we mean that only $C$ and $M$ get to know the value,
while any third party can not even tell that a signal was exchanged.
It is then trivial to build a {\em redemption protocol} on top.
Note that signaling does not require that the value can be arbitrarily chosen by $C$.
We will provide a signaling protocol specifically and only for bitcoin.
Both parts are depicted in Figure~\ref{fig:anon}. 

\subsection{Signaling protocol}

While it is unfeasible for $M$ to monitor the whole distributed filesystem, 
we assume it is feasible for $M$ to monitor the blockchain.
The rationale is that the blockchain is small compared to a distributed filesystem because of bitcoin's transaction fees.
In fact, $M$ does this kind of monitoring in any of the before-mentioned protocols, to detect incoming payments.
Therefore, we carry out the signaling protocol over the blockchain.
$M$ will be required to do work comparable to a full node.

A specially crafted transaction serves as the {\em signal}.
The requirement is that an observing third party cannot distinguish the signaling transaction from a regular transaction,
in particular 
i) cannot link the signal to $M$,
and
ii) cannot read off the signalled value.
The underlying idea is to carry out a Diffie-Hellman key exchange between $M$'s reputation key $P$ and a pubkey that appears in the transaction for which $C$ holds the private key.
It does not matter whether this second pubkey appears in an input or output of the transaction.

As before, we assume the reputation key $P$ is stored on $C$'s trusted path device $T$. 
Then $C$ carries out the following steps on $T$:
\begin{enumerate}
\item build an arbitrary transaction $X$,
leaving an arbitrary amount for an output that is to be added,
\item let $a$ be any pubkey appearing in $X$, either in an input or in an output to a pubkey,
\item let $s$ be the private key corresponding to $a$,
\item let $c$ be the $x$-coordinate of the curve point $P^{s}$,
\item use $d_{\text{addr}}(P,c)$ as an output address of $X$.
\end{enumerate}
The signalled value is $c$.
The curve point $P^{s}=a^{K}$ is the result of a Diffie-Hellman key agreement between $a$ and $P$.
To detect the signal, $M$ carries out the following steps for each pubkey $a$ that appears on the blockchain:
\begin{enumerate}
\item let $c$ be the $x$-coordinate of the curve point $a^{K}$,
\item check whether $d_{\text{addr}}(P,c)$ is an output address of $X$, if so $c$ is the signalled value.
\end{enumerate}

In case $M$ denies to have received the signal, $C$ reveals the curve point $P^{s}$.
This allows anyone to find the signaling transaction by searching for the address $d_{\text{addr}}(P,c)$.
However, it remains to prove that a given curve point is indeed equal to $P^{s}$ without revealing $s$.
This is possible by providing a certain type of signature~\cite{chaum-pedersen},
which is based on a zero-knowledge proof 
that two discrete logarithms with respect to different bases are equal. 
The signing key is $s$ and the signed message is $P$.
The resulting signature ``proves'' that the discrete logarithm of $P^{s}$ with respect to basis $P$
equals the discrete logarithm of $a$ with respect to basis $g$
(they are both $s$).
$C$ carries out the following steps to produce the signature $\sigma$
(cf.\ section 3.2~of~\cite{chaum-pedersen}).
The symbol $\|$ means concatenation of strings.
\begin{enumerate}
\item choose a random number $u$,
\item let $v=H(P\|P^s\|g^u\|P^u)$, 
\item let $\sigma=(P^s,g^u,P^u,u+vs)$.
\end{enumerate}
A third party verifies $\sigma$ by checking these three equations
(cf.\ section 3.2~of~\cite{chaum-pedersen}):
\[ v=H(P\|P^s\|g^u\|P^u),\quad g^{u+vs}=g^ua^v, \quad P^{u+vs}=P^u(P^s)^v. \]
Note that the signalled value is unique for the pair $(P,a)$.
This implies that $a$ cannot be used twice for signaling to the same $P$.

At the end of the protocol,
$M$ has control over the arbitrary amount that was sent to the output $d_{\text{addr}}(P,c)$.
As an additional but unused feature,
if that output gets redeemed on the blockchain
then $C$ has a proof that his signal was received without revealing $P^{s}$.

\subsection{Redemption protocol}

To redeem a contract $x$, $C$ carries out the following steps: 
\begin{enumerate}
\item signal a unique value $c$ to $M$,
\item derive a symmetric key $c'$ from $c$, e.g.\ $c'=c$ or $c'=H(c)$,
\item encrypt $x$ symmetrically with key $c'$ to $x'$,
\item post data $x'$ under the filename $H(c')$ to the distributed filesystem.
\end{enumerate}

When receiving a signal with value $c$, 
$M$ carries out the following steps to retrieve the contract:
\begin{enumerate}
\item derive the symmetric key $c'$ from $c$,
\item retrieve file $x'$ with filename $H(c')$ 
\item decrypts $x'$ symmetrically with key $c'$ to $x$
\end{enumerate}
Redemption is complete,
next $M$ looks for a transaction with output $d_{\text{addr}}(P,x)$ on the blockchain.

The signaling and paying protocol can be combined into one transaction with two outputs.
Since both outputs can be redeemed by $M$, the funds can be distributed arbitrarily between the two outputs.

\subsection{Signaling variations}

A variation is to use $d_{\text{addr}}(a,c)$ as an output address instead of $d_{\text{addr}}(P,c)$. 
Then $C$ remains in control of the arbitrary amount sent there. 
However, this has some security implications:
the private keys of $a$ and $d_{\text{addr}}(a,c)$ can be computed from each other by anyone knowing $c$, e.g.\ by $M$.
Therefore, after giving the signal, $C$ should never separate the private keys of $a$ and $d_{\text{addr}}(a,c)$,
but always keep them under the exact same security level.

\section{Conclusions}

We have introduced a payment protocol for customer-merchant relations.
Merchants are identified by a public key which is their pseudonym. 
The pseudonym not only founds the reputation of the merchant,
but it also specifies where funds addressed to the merchant are to be sent.

No authenticated or encrypted communication is required between customer and merchant,
and no signatures are generated by the merchant at order time.
Consequently, the protocol can work on compromised merchant infrastructure. 
The only requirement is that the customer possesses a trusted path device,
on which all sensitive parts of the protocol are concentrated.
A one-way communication from the payer to the merchant is required after the payment,
but again security does not demand it to be encrypted or authenticated.

The customer's payment transaction simultaneously serves as a receipt for the customer.
It proves what has been paid for and who received the funds.

An extension of the protocol allows that all communication between customer and merchant is carried out 
openly via a distributed filesystem.
In this way the merchant can hide all his infrastructure from the customer and stay truly pseudonymous.

The protocol is suitable for bitcoin as the underlying payment system.
Since bitcoin transactions are publicly recorded,
the protocol allows to exchange funds and receipt in an atomic transaction.
Despite this public record, transactions cannot be linked to the merchant pseudonym by third--party observers.

An implementation with bitcoin would require little effort.
One requirement is the standardization of the ``labeled wallets'' that we introduced,
and their implementation across all bitcoin clients. 
A second requirement is an authentication infrastructure for the merchant's pseudonyms,
preferably inside the bitcoin client.
Since pseudonyms are equivalent to bitcoin addresses,
the customer can simply tag those bitcoin addresses
which are pseudonyms of merchants that he has started a business relation with.
To use all features of the protocol, however, pseudonyms are to be distributed via a public key infrastructure or web-of-trust.

Hardware wallets, which are currently under development, can serve as the trusted path device for the customer.
The requirement is that a bitcoin client with the above mentioned additional features is running on the hardware wallet.
Also the hardware wallet is required to be able to display a considerable amount of text.

\subsection*{Acknowledgements}

The forum on bitcointalk.org is a vast source of information on all bitcoin-related topics,
including ideas regarding payment protocols, customer-merchant best practices, and feature extensions based on elliptic curve arithmetic.
While it was impossible for us to browse through all forum threads,
we point out a few references that we have come across and that seem particularly close to core ideas of our proposal.
The first mention known to us of customer created payment addresses is by Gavin Andresen~\cite{andresen}.
The self-identification feature of the signaling protocol
via a Diffie-Hellman key exchange was invented by ByteCoin~\cite{bytecoin2}.
A payment protocol involving structured payment descriptors has been suggested by Pieter Wuille (sipa)~\cite{wuille}.
We thank all contributors on bitcointalk.org that we forgot to mention here.

Sergio Lerner~\cite{lerner} outlines a payment protocol very similar to ours, including offline redemption.
We became aware of his post only at the final stages of the preparation of this manuscript
and hope to have expanded on his ideas substantially.


\end{document}